\documentclass[prd,aps,twocolumn,superscriptaddress,floatfix]{revtex4}
\usepackage{graphicx,natbib,times}
\usepackage{url}
\usepackage{color}
\usepackage{rotating}
\usepackage{hyperref}
\usepackage[normalem]{ulem}
\hypersetup{
    colorlinks=true,
   linkcolor=red,
    citecolor=blue,
}

\def\etal{{\frenchspacing\it et al.}}

\def\beq{\begin{equation}}
\def\eeq{\end{equation}}
\def\ben{\begin{eqnarray}}
\def\een{\end{eqnarray}}
\def\numass{M_{\nu}}
\def\nus{\texttt{MassiveNuS}}

\def\nulcdm{\nu\Lambda{\rm CDM}}
\def\ev{\,{\rm eV}}
\def\bx{{\bf x}}
\def\br{{\bf r}}
\def\ez{\eta_{0.0}(r)}
\def\el{\eta_{0.1}(r)}
\def\eh{\eta_{0.6}(r)}
\def\be{\hat{\bf e}}
\def\sig8{\sigma_{8}}
\def\munit{\,h^{-1}M_{\odot}}
\def\dunit{\,h^{-1}{\rm Mpc}}
\begin{document}
\title{Weighing the Neutrinos with the Galaxy Shape-Shape Correlations}

\author{Jounghun Lee}
\email{jounghun@astro.snu.ac.kr}
\affiliation{Department of Physics and Astronomy, Seoul National University, Seoul 08826, Republic of Korea}
\author{Suho Ryu}
\email{ryu@snu.ac.kr}
\affiliation{Department of Physics and Astronomy, Seoul National University, Seoul 08826, Republic of Korea}
\begin{abstract}
The galaxies form and evolve in the early epochs through the anisotropic merging process along the primary narrow filaments, in the direction of 
which their shapes become elongated and intrinsically aligned. The nonlinear evolution of the cosmic web broadens the primary filaments, by 
entangling them with multiple secondary filaments, which has an effect of reducing the anisotropy of the merging process and in consequence 
weakens the galaxy shape-shape correlations in the later epochs.  
Assuming that the degree of the nonlinearity and complexity of the cosmic web depends on the nature of dark matter,  we propose a hypothesis 
that the galaxy shape-shape correlation function, $\eta(r)$, may be a powerful complimentary probe of the total neutrino mass, $\numass$. 
Testing this hypothesis against a high resolution N-body simulation, we show that the $\numass$-dependence of $\eta(r)$ at $z=0$ is 
sensitive enough to distinguish between $\numass=0.0\ev$ and $\numass=0.1\ev$.  We also show that the differences 
in $\eta(r)$ at $r\le 5\dunit$ between the models with massless and massive neutrinos cannot be explained by their differences in the 
small-scale density powers, $\sig8$, which implies that the galaxy shape-shape correlation function has a potential to break the notorious 
cosmic degeneracy between $\numass$ and $\sig8$. 

\end{abstract}

\pacs{ }

\maketitle

{\it Introduction.} 
The shapes of the Milky way-sized galaxies in the universe are observed to be mutually and intrinsically cross-correlated \cite{obs_eta}.  
The recent numerical studies based on high-resolution cosmological simulations have suggested that the galaxy shape-shape correlations  
should be closely linked with the anisotropic merging process in the filamentary cosmic web, through which the galactic halos form and 
evolve \cite{vera11}.  In the early epochs, the galactic halos become elongated in the directions of the primary narrow filaments along 
which the merging events most frequently occur, and in consequence their shapes become mutually correlated over the scales comparable 
to the spatial extents of the primary filaments \cite{vera11,anisotropy}.

In the later epochs, however, the cosmic web develops a more complex structure where the primary filaments entangled with 
multiple secondary filaments become broader than the sizes of the galactic halos \cite{vera11}. In this nonlinearly evolved cosmic web, the 
merging process would occur more or less isotropically \cite{zomg1}, which has an effect of diminishing the strength of the halo shape-shape 
correlations.  
Given that the presence of the cosmic web is caused by the large-scale coherence of the tidal shear fields \cite{cosmicweb}, whose 
nonlinear evolution can be retarded by the free streaming of hot dark matter particles like massive neutrinos \cite{nu_cosmos,nu_web}, 
the degree of the nonlinearity and complexity of the cosmic web is expected depend on the nature of dark matter. 

We claim here that the halo shape-shape correlations can be a new complementary probe of the relic neutrinos, which must possess non-zero 
mass according to the results of the solar neutrino oscillation experiments \cite[see][for a review]{nu_mass}. 
Our goal here is to numerically prove this claim by utilizing the data from recently available high-resolution $N$-body simulations performed for 
the $\nulcdm$ cosmologies (the cosmological constant $\Lambda$ + cold dark matter CDM + three species of neutrinos $\nu$ with the total 
mass, $\numass\ge 0.0\ev$). 
 
{\it Numerical Analysis and Results.} 
The halo shape-shape correlation function, $\eta(r)$, is defined as \cite{lee-etal08} 
\beq
\label{eqn:eta}
\eta(r)\equiv \langle\vert\be({\bx})\cdot\be({\bx}+{\br})\vert^{2}\rangle - \frac{1}{3}\, ,
\eeq
where $\be({\bx})$ and $\be({\bx}+{\br})$ denote the unit vectors in the major principal directions of the inertia momentum tensors 
of the DM halos located at the positions of $\bx$ and ${\bx}+{\br}$, respectively. 
From here on, we call, $\be$, {\it a shape vector of a DM halo}.
The first term in the right-hand side of Equation (\ref{eqn:eta}) represents the ensemble average of the squares of the inner products  
between $\be({\bx})$ and $\be({\bx}+{\br})$, which is $1/3$ if there is no correlation.  

To investigate the $\numass$-dependence of $\eta(r)$, we utilize the publicly available data from the Cosmological Massive Neutrino 
Simulations ($\nus$) conducted by \cite{nus} for a number of $\nulcdm$ models with diverse initial conditions. 
Performed in a periodic box of a side length $512\dunit$,  the $\nus$ has mass and particle resolutions as high as $10^{10}\munit$ and 
$1024^{3}$, respectively. 
In the $\nus$, the analytic linear response approximation was adopted to track down the positions and velocities of the relic neutrinos, 
while the Rockstar algorithm \cite{rockstar} was applied to the phase space distributions of the DM particles to find the distinct halos and 
their subhalos as well.  The Rockstar catalog from the $\nus$ provides information on various properties of each object including $\be$, 
its position $\bx=(x_{i})$, minor-to-major axial ratio $S$, virial mass  $M_{h}$ and radius $r_{h}$, and scale radius $r_{s}$. 

For the current scrutiny on the $\numass$-dependence of $\eta(r)$, we consider only three $\nulcdm$ models with massless, light and 
heavy neutrinos (corresponding to $\numass=0.0,\ 0.1$ and $0.6 \ev$, respectively).  For the three models, the key cosmological parameters, 
other than $\numass$, such as the matter density parameter, baryon density parameter, amplitude of the primordial power spectrum, 
spectral index and dimensionless Hubble parameter are identically set at $\Omega_{m}=0.3$, $\Omega_{b}=0.047$, $A_{s}=10^{-9}$, 
$n_{s}=0.97$ and $h=0.7$, respectively. Given these initial conditions, the values of the rms fluctuations of the linear density contrasts 
within a spherical radius $8\dunit$ were evaluated to be $0.85$, $0.83$ and $0.74$ for the $\nulcdm$ models with massless, light and heavy 
neutrinos, respectively.  

For each model, we make a sample of the well resolved distinct halos by eliminating the subhalos from the catalog of the Rockstar objects 
at $z=0$ and selecting only those halos with $M_{h}\ge 10^{12}\munit$ containing $100$ or more DM particles. For each pair of the distinct 
halos in the sample, we measure their separation distance $r$, and compute the square of the inner product between their shape vectors.  
Subtracting $1/3$ from its spatial average taken over those pairs of the distinct halos with $r$ in the differential bin $[r, r+dr]$, we numerically 
determine $\eta(r)$. The ensemble average over the realizations in Equation (\ref{eqn:eta}) is replaced by the spatial average over $\bx$, which 
can be justified by the ergodic theorem.  To estimate the errors in the determination of $\eta(r)$ at a given $r$-bin to which $n_{\rm pair}$ 
pairs of the distinct halos belong, we generate $n_{\rm pair}$ sets of $10^{6}$ random unit vectors and determine the one standard 
deviation scatter, $\sigma_{\eta}$, among the sets, as done in \cite{scatter}. Finally, the errors in $\eta(r)$ at each $r$-bin is estimated to be  
$\sigma_{\eta}/\sqrt{n_{\rm pair}}$.

\begin{figure}[ht]
\resizebox{\columnwidth}{!}{\includegraphics{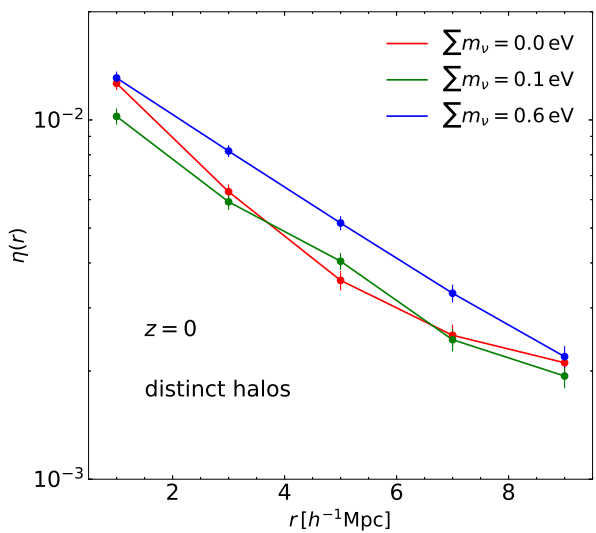}}
\caption{Intrinsic shape-shape correlation function of the distinct DM halos from the $\nus$ at $z=0$ for three different 
$\nulcdm$ models.
\label{fig:distinct} }
\end{figure}
Figure \ref{fig:distinct} plots the numerically obtained $\eta(r)$ in the distance range of $r< 10\dunit$ from the samples of the distinct halos 
at $z=0$ for the three $\nulcdm$ models.  The results at longer distances $r\ge 10\dunit$ are found to carry large uncertainties and thus 
left out.  From here on, we let $\ez,\ \el$ and $\eh$ denote the halo shape-shape correlation functions from the 
models with massless, light and heavy neutrinos, respectively. 

As can be seen, $\ez$ shows a distinctly different behavior compared with $\el$ and $\eh$. 
For the latter two functions, their $\numass$-dependence shows a simple pattern, $\el>\eh$ in the whole range of $r$, which is consistent 
with our expectation based on the following logic. In the model with heavy neutrinos that has a lower value of $\sig8$ than the model with light 
neutrinos, the shape-shape correlations of the galactic halos retain better the initial strengths that they acquired in the early epochs through 
the preferential merger events along the narrow primary filaments.  

Meanwhile, the difference between $\ez$ and the other two functions shows a more complicated pattern. 
At $3\le r/[\dunit]<8$,  we find $\ez\sim\el < \eh$, which can be ascribed to the differences in $\sig8$ among the three models. 
However, at $r<3\dunit$,  we witness a much more rapid increase of $\ez$ with the decrement of $r$ than $\el$ and $\eh$, which 
leads $\el<\ez\sim \eh$ at $r\sim 1\dunit$. This behavior cannot be explained by the aforementioned simple logic, since the two models with 
massless and heavy neutrinos have widely different values of $\sig8$.  Some other mechanism like the gravitational halo-halo interactions 
must counteract the effect of the high value of $\sig8$ to enhance $\ez$ at $r\le 2\dunit$ only for the case of massless neutrinos.  

In practice, what is more readily observable is not the correlations between the shapes of the distinct halos but the correlations between 
the projected shapes of the galaxies. We make a sample of the galactic halos in the mass range, $10^{12}\le M/[\munit]<10^{13}$, 
from the $\nus$ Rockstar catalog at $z=0$ and determine $\eta(r)$ from this sample by repeating the whole process described in the above. 
Note that the sample of the galactic halos includes the subhalos embedded in larger distinct halos as well as the distinct DM halos 
with no subhalos. The former correspond to the cluster/group galaxies while the latter to the field galaxies. 
\begin{figure}[ht]
\resizebox{\columnwidth}{!}{\includegraphics{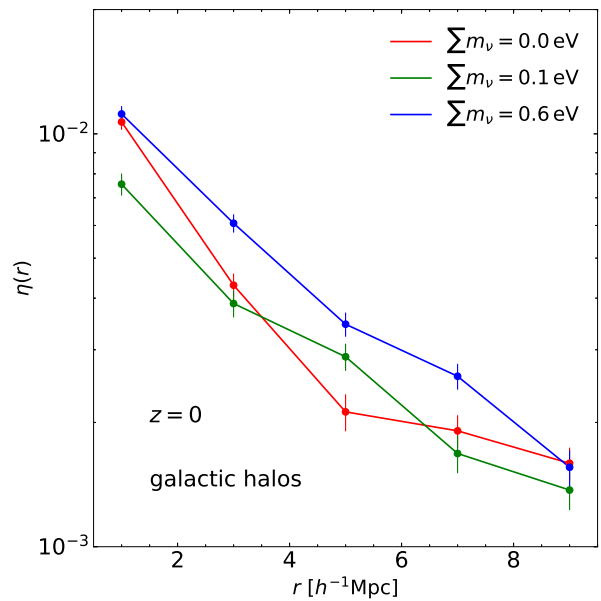}}
\caption{Same as Figure \ref{fig:distinct} but for the galactic halos in the mass range of $1\le M_{h}/(10^{12}\munit)<10$.
\label{fig:galactic} }
\end{figure}

Figure \ref{fig:galactic} plots $\ez,\ \el$ and $\eh$ from the sample of the galactic halos. As can be seen, the results from the galactic 
halos are similar to those from the distinct halos shown in Figure \ref{fig:distinct}. We find $\el\ll \eh$ in the whole range of $r$ and 
$\el\ll\ez\sim\eh$ at $r\sim 1\dunit$. A notable difference between the results shown in Figures \ref{fig:distinct} and \ref{fig:galactic} 
is that $\ez$ changes its rate, $d\ez/dr$, more abruptly around $r\sim 5\dunit$, for the case of the galactic halos. It increases with 
the decrement of $r$ much more mildly at $r>5\dunit$ and much more rapidly at $r\le 5\dunit$ than $\el$ and $\eh$. 
Note that $\ez$ is substantially lower than $\el$ at $r\sim 5\dunit$ while it is significantly higher than $\el$ at $r\sim 1\dunit$. 

\begin{figure}[ht]
\resizebox{\columnwidth}{!}{\includegraphics{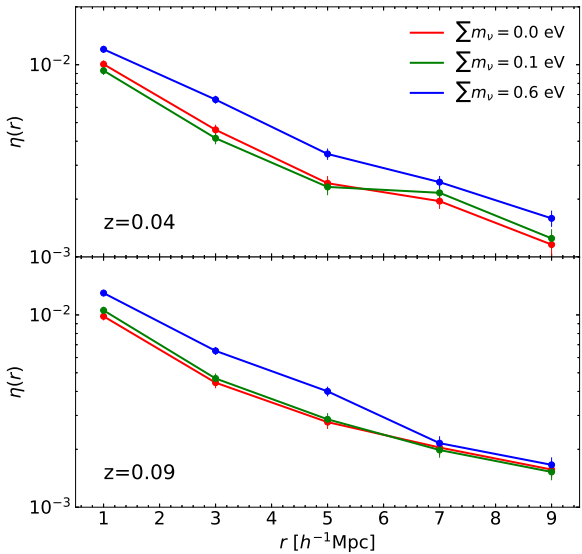}}
\caption{Same as Figure \ref{fig:galactic} but at $z=0.04$ (top panel) and at $z=0.09$ (bottom panel).
\label{fig:highz} }
\end{figure}
We also investigate how $\ez,\ \el$ and $\eh$ from the galactic halos evolve with redshifts, by conducting the same analysis at two higher 
redshifts, $z=0.04$ and $0.09$,  the results of which are shown in the top and bottom panels of Figure \ref{fig:highz}, respectively. 
As can be seen, we find $\ez\sim\el\ll \eh$ in the whole range of $r$ at both of the higher redshifts, which trends are well 
explained by the differences in $\sig8$ among the three models. Note also that $\ez$ is closer to $\el$ in the whole range of $r$ 
at $z=0.09$ than at $z=0.04$. This result indicates that the mechanism responsible for the significant deviation of $\ez$ from $\el$ 
in the range of $r\le 5\dunit$ at $z=0$ should operate much less effectively at higher redshifts. 

\begin{figure}[ht]
\resizebox{\columnwidth}{!}{\includegraphics{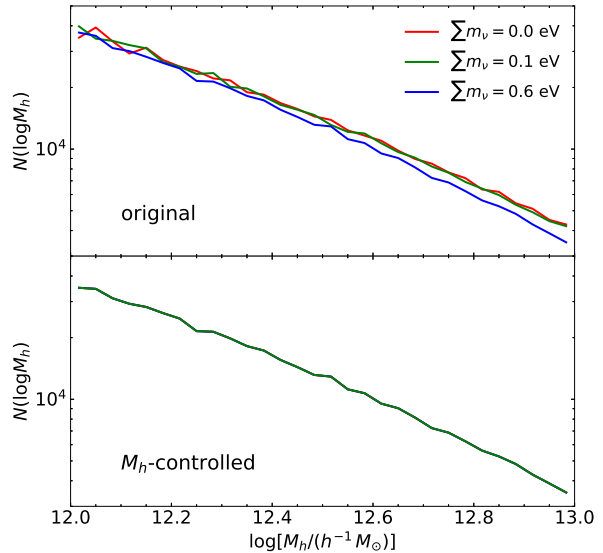}}
\caption{Number distributions of the galactic halos at $z=0$ from the original (top panel) and mass-controlled (bottom panel) 
samples for the three $\nulcdm$ models.
\label{fig:mdis}}
\end{figure}

Since the strength of the halo shape-shape correlations has been known to depend on the physical properties of the halos such as their 
masses, sphericities and formation epochs \cite{dep}, we would like to see whether or not the differences in $\eta(r)$ among the three $\nulcdm$ 
models are due to their differences in the distributions of the halo properties. 
Splitting the logarithmic mass range, $12\le \log M_{h} < 13$ into multiple differential bins and counting the numbers of the 
galactic halos whose logarithmic masses fall in each bin for each of the three $\nulcdm$ models , we determine the number distributions, 
$n(\log M_{h})$, of the galactic halos as a function of $\log M_{h}$ at $z=0$, the results of which are shown in the top panel of 
Figure \ref{fig:mdis}.  As expected, the model with heavy neutrinos exhibits the lowest amplitude of $n(\log M_{h})$, while the other 
two models yield quite similar mass distributions. 

\begin{figure}[ht]
\resizebox{\columnwidth}{!}{\includegraphics{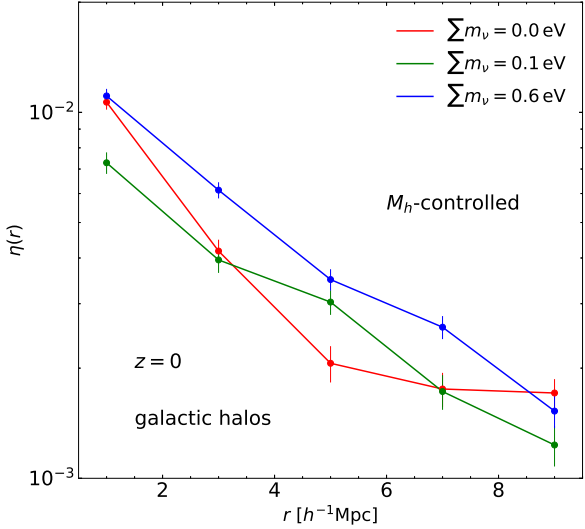}}
\caption{Same as Figure \ref{fig:galactic} but from the $M_{h}$-controlled samples.
\label{fig:mcon} }
\end{figure}

Let $n_{0.0}(\log M_{h}),\ n_{0.1}(\log M_{h})$ and $n_{0.6}(\log M_{h})$ denote the mass distributions of the galactic halos for 
the models with massless, light and heavy neutrinos, respectively. Defining $n_{\rm min}$ as 
$n_{\rm min}\equiv \min\{n_{0.0},\ n_{0.1},\ n_{0.6}\}$, we select $n_{\rm min}$ galactic halos at each mass bin 
from each model to create three controlled samples of the galactic halos that have identical mass distributions, which 
are shown in the bottom panel of Figure \ref{fig:mdis}. Using these three controlled samples, we refollow the whole procedure to 
redetermine $\ez,\ \el$ and $\eh$ at $z=0$ and show the results in Figure \ref{fig:mcon}. As can be seen, no appreciable difference 
is found between the results from the original and controlled samples. Even though the three controlled samples have the identical 
mass distributions, they still exhibit significant differences in $\eta(r)$, which indicates that the differences in $n(\log M_{h})$ are not 
responsible for the differences among $\ez,\ \el,\ \eh$. 

\begin{figure}[ht]
\resizebox{\columnwidth}{!}{\includegraphics{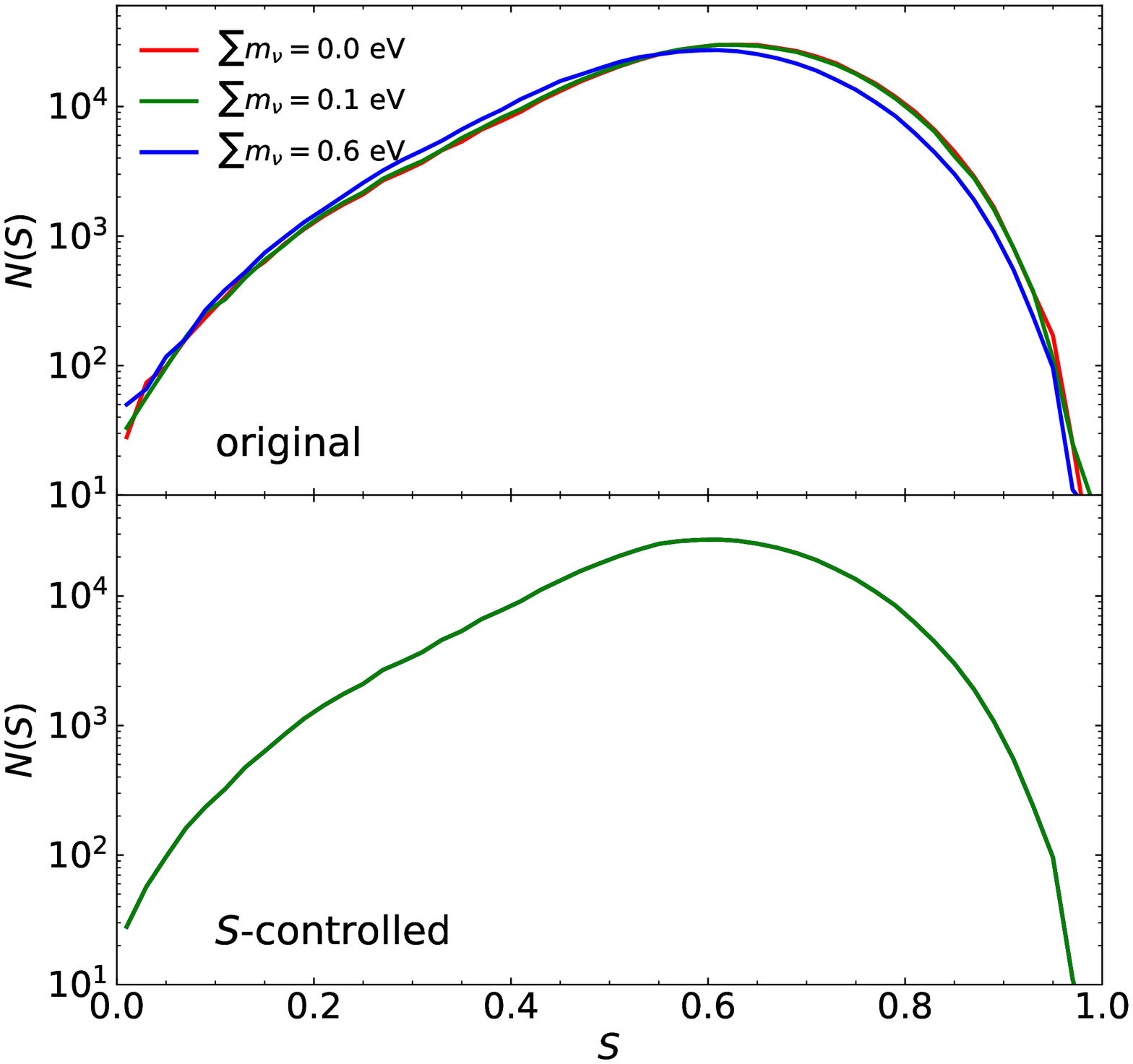}}
\caption{Number distributions of the galactic halos at $z=0$ from the original (top panel) and sphericity-controlled (bottom panel) 
samples for the three $\nulcdm$ models.
\label{fig:sdis}}
\end{figure}

\begin{figure}[ht]
\resizebox{\columnwidth}{!}{\includegraphics{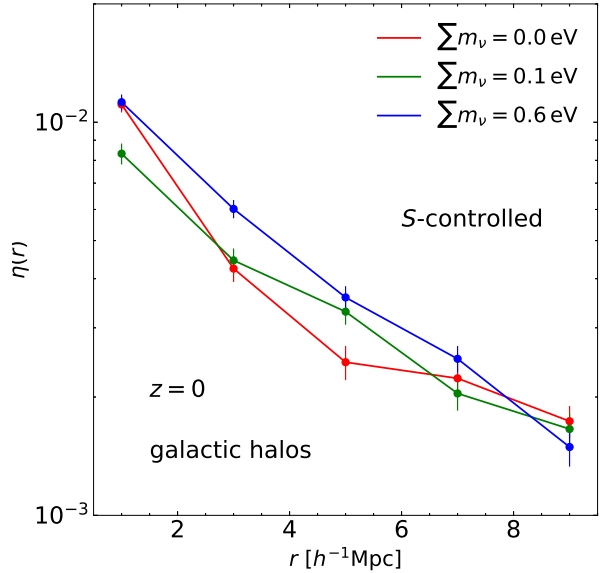}}
\caption{Same as Figure \ref{fig:galactic} but from the $S$-controlled samples.
\label{fig:scon} }
\end{figure}

In a similar manner, we also create three controlled samples of the galactic halos, which have the identical distributions 
of the sphericity, $S$, of the galactic halos, where $S$ is defined as the minor to major axial ratio of a galactic halo \citep{shape}. 
Figure \ref{fig:sdis} plots the $S$-distributions from the original and controlled samples of the galactic halos for the three $\nulcdm$ 
models at $z=0$.  
As can be seen in the top panel, the galactic halos in the model with heavy neutrinos tend to be more aspherical than those in the 
other two models. Figure \ref{fig:scon} plots $\ez,\ \el,\ \eh$ from the $S$-controlled samples, revealing that although differences between 
$\el$ and $\eh$ in the whole range of $r$ is reduced compared with the results from the original sample, the controlled samples still 
exhibit the same degree of the key differences between $\ez$ and $\el$ at $r\sim 1\dunit$ as well as between $\ez$ and $\eh$ at 
$r\sim 3\dunit$. This result implies that the $\numass$-dependence of $\eta(r)$ cannot be ascribed to the differences in the $S$-distributions 
among the three models. 


\begin{figure}[ht]
\resizebox{\columnwidth}{!}{\includegraphics{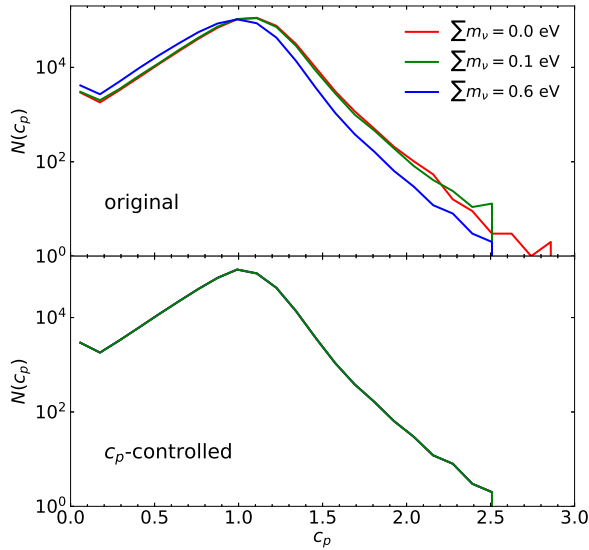}}
\caption{Number distributions of the galactic halos at $z=0$ from the original (top panel) and concentration parameter-controlled 
(bottom panel) samples for the three $\nulcdm$ models.
\label{fig:cdis}}
\end{figure}

\begin{figure}[ht]
\resizebox{\columnwidth}{!}{\includegraphics{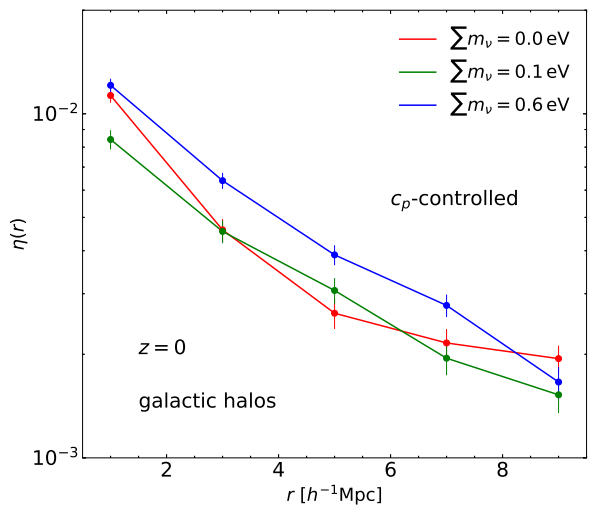}}
\caption{Same as Figure \ref{fig:galactic} but from the $c_{p}$-controlled samples.
\label{fig:ccon} }
\end{figure}

Similarly, we also examine if and how  the differences in the formation epochs of the galactic halos, $c_{p}$, among the three models 
contribute to the differences in $\ez,\ \el,\ \eh$, where the formation epoch of a galactic halo is determined as $c_{p}\equiv r_{s}/r_{v}$. 
The $c_{p}$-distributions from the original and controlled samples for the three models are shown in Figure \ref{fig:cdis}, which reveals 
that the galactic halos in the model with heavy neutrinos tend to have lower formation epochs than those in the other two models, as expected. 
The results of $\ez,\ \el,\ \eh$ from the three $c_{p}$-controlled samples are depicted in Figure \ref{fig:ccon}. Note that while the difference 
between $\ez$ and $\el$ at $r\sim 5\dunit$ is substantially reduced, the difference between $\ez$ and $\el$ at $r\sim 1\dunit$ as well as that 
between $\ez$ and $\eh$ at $r\sim 3\dunit$ is still quite significant, even when the $c_{p}$-controlled samples are used. 
This result implies that the differences in the $c_{p}$-distributions among the three models have little to do with the differences 
between $\ez$ and $\el$ at $r\sim 1\dunit$ nor between $\ez$ and $\eh$ at $r\sim 3\dunit$, while they should be largely responsible 
for the differences between $\ez$ and $\el$ at $r\sim 5\dunit$. 

\begin{figure}[ht]
\resizebox{\columnwidth}{!}{\includegraphics{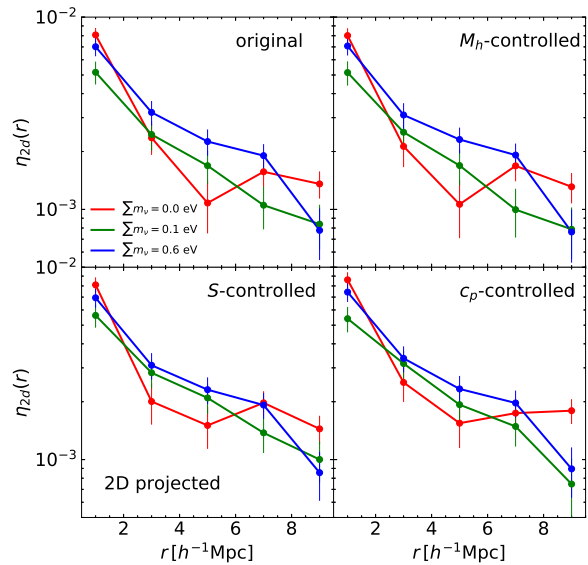}}
\caption{Same as Figure \ref{fig:galactic} but for the case that the shape vectors are projected onto the 2D plane. 
\label{fig:2d}}
\end{figure}

Now that we have numerically found a robust evidence for the $\numass$-dependence of $\eta(r)$ that cannot be attributed to its 
$\sig8$-dependence, we would like to assess how detectable the signal is in practice.  Regarding the direction of the $\hat{x}_{3}$-axis 
as a direction of the line of sight normal to the flat plane of the sky, we project $\be$ onto the $\hat{x}_{1}$-$\hat{x}_{2}$ plane and 
renormalize it to obtain a two dimensional unit shape vector, $\be_{2d}$.  Then, we determine the two-dimensional shape-shape correlation 
function as $\eta_{2d}(r)\equiv \langle\vert\be_{2}({\bf x})\cdot\be_{2d}({\bf x}+{\bf r})\vert^{2}\rangle - 1/2$ for each of the three 
$\nulcdm$ models. The four panels of Figure \ref{fig:2d} plots $\eta_{2d}(r)$ from the original, $M_{h}$-controlled, $S$-controlled 
and $c_{p}$ controlled samples for the three models.  Although $\eta_{2d}(r)$ carries large errors compared with $\eta(r)$ in the whole 
range of $r$, the differences between $\eta_{2d,0.0}$ and $\eta_{2d,0.1}$ at $r\sim 1\dunit$ seem to remain significant for all of the four cases.

To assess more rigorously the statistical significance, we perform the Kolmogorov–Smirnov (KS) test of the null hypothesis that there is no 
difference between $\eta_{2d,0.0}$ and $\eta_{2d,0.1}$ at $r\sim 1\dunit$. For this KS test, we first determine the cumulative probability 
distribution, $P(<\cos\theta_{2d})$, with $\cos\theta_{2d}\equiv \vert\be_{2d}({\bf x})\cdot\be_{2d}({\bf x}+{\bf r})\vert$ with 
$\vert{\bf r}\vert=1\dunit$ for each sample. 
If $\be_{2d}$ is completely random, then $P(<\cos\theta_{2d})=\cos\theta_{2d}$. If $\be_{2d}$ is cross-correlated, then 
$P(<\cos\theta_{2d})<\cos\theta_{2d}$ in the range of $0<\cos\theta_{2d}<1$. 
The stronger the cross-correlations are, the smaller $P(<\cos\theta_{2d})$ is than $\cos\theta_{2d}$.
We calculate the KS statistics as $\max \vert P_{0.0}-P_{0.1}\vert$, where $P_{0.0}$ and $P_{0.1}$ represent $P(<\cos\theta_{2d})$ 
from the models with massless and light neutrinos, respectively, and evaluate the confidence level at which the null hypothesis is 
rejected with this KS statistics. 

\begin{figure}[ht]
\resizebox{\columnwidth}{!}{\includegraphics{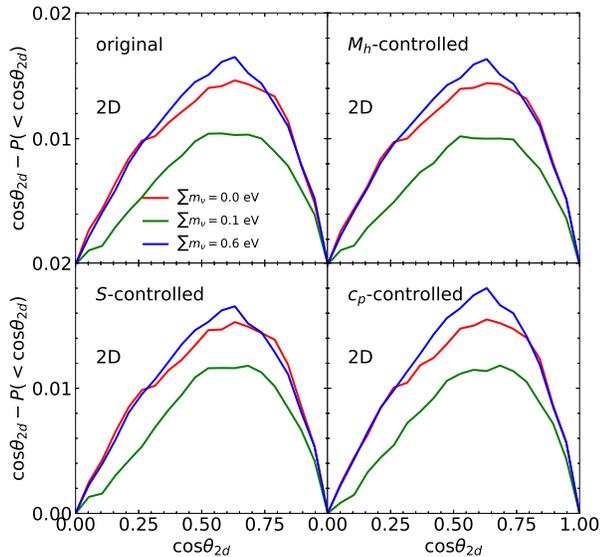}}
\caption{Differences between $\cos\theta_{2d}$ and $P(<\cos\theta_{2d})$ from the original, $M_{h}$-controlled, 
$S$-controlled and $c_{p}$-controlled samples of the galactic halos for the three $\nulcdm$ models.
\label{fig:kstest} }
\end{figure}

Figure \ref{fig:kstest} plots $\cos\theta_{2d}-P(<\cos\theta_{2d})$ obtained from the pairs of the galactic halos with separation distance 
$r\sim 1\dunit$ belonging to the original, $M_{h}$-controlled, $S$-controlled and $c_{p}$ controlled samples for the two $\nulcdm$ 
models with massless and light neutrinos. As can be seen, the KS statistics are large enough to reject the null hypothesis 
at the confidence level higher than $99.9\%$ for all of the four cases. In other words, the two models with massless and light neutrinos 
significantly differ in the strengths of the cross-correlations of the projected shapes of the galactic halos at $r\sim 1\dunit$. 
It is worth mentioning that the statistical significance of the differences between $\ez$ and $\eh$ at $3\dunit$, however, seems to be severely 
reduced by the projection effects. A larger sample of the galaxies should be required to beat down the projection effects for the detection 
of this signal in practice.

{\it Discussions and Conclusion.} 
We have numerically determined the galaxy shape-shape correlation functions, $\eta(r)$, by analyzing the data from the $\nus$ 
for three different $\nulcdm$ models with $\numass=0.0,\ 0.1,\ 0.6\ev$. It has been shown that the three models significantly 
differ in $\eta(r)$ at $r\le 5\dunit$ among one another, which implies that $\eta(r)$ should be in principle a powerful new probe of 
$\numass$. 

The robustness of this new probe has been tested against controlling the properties of the galactic halos that are correlated with 
the strengths of $\eta(r)$. Even when the correlations are measured from the projected shapes of the galactic halos in the two 
dimensional space, the differences among the three models at $r\le 5\dunit$ have been found to statistically significant. 

The galaxy shape-shape correlation has two advantages as a complementary probe of $\numass$. First, it is sensitive enough to 
distinguish between $\numass=0.0$ and $0.1\ev$. Second, it has a potential to break the $\sig8$-$\numass$ degeneracy since 
the differences in $\eta(r)$ among the three models cannot be explained by the differences in $\sig8$.

A backup work, however, has to be done before making a practical use of $\eta(r)$ as a probe of $\numass$.  
For a direct comparison with observational data, it will be necessary to determine the shape vectors of the galaxies and their cross-correlations 
from their baryonic gas particles, for which the data from the cosmological hydrodynamic simulations for the $\nulcdm$ models will be required. 
It will be also necessary to take into proper account the effects of the misalignments between the DM and baryonic particle distributions 
as well as the feedbacks of the non-gravitational processes, since they could be strong enough to alter the behaviors of $\eta(r)$ \cite{baryon}.  
Our future work will be in this direction.

{\it Acknowledgements.} 
We thank the Columbia Lensing group for making their suite of simulated maps available at the website (http://columbialensing.org), and 
NSF for supporting 
the creation of those maps through grant AST-1210877 and XSEDE allocation AST-140041. We thank the New Mexico State University (USA) and Instituto de 
Astrofisica de Andalucia CSIC (Spain) for hosting the Skies \& Universes site for cosmological simulation products. 
We acknowledge the support by Basic Science Research Program through the National Research Foundation (NRF) of Korea 
funded by the Ministry of Education (No.2019R1A2C1083855).

\end{document}